# Wave-Particle Duality and the Coherent Quantum Domain Picture

Alan M. Kadin[*]
Princeton Junction, NJ 08550 USA
March 7, 2006

**Abstract:**
It is proposed that the paradox of wave-particle duality in quantum mechanics may be resolved using a physical picture analogous to magnetic domains. Within this picture, a quantum particle represents a coherent region of a quantum wave with characteristic total energy, momentum, and spin. The dynamics of such a state are described by the usual linear quantum wave equations. But the coherence is maintained by a nonlinear self-interaction term that is evident only during transitions from one quantum state to another. This is analogous to the self-organizing property of domains in a ferromagnetic material, in which a single domain may appear as a stable macro-particle, but with rapid transitions between different domain configurations also possible. For the quantum case, this implies that the "collapse of the wave function" is a real dynamical physical process that occurs continuously in spacetime. This picture may also permit the resolution of apparent paradoxes associated with quantum measurement and entangled states.

## I. Introduction

Is light composed of particles or waves? That question, going back hundreds of years, appeared to be definitively settled by Maxwell's equations in the mid-19[th] century, in favor of waves. However, by the beginning of the 20[th] century, it had become clear that phenomena such as black-body radiation and the photoelectric effect required the existence of particles of light known as photons. Furthermore, it was shown that particles of matter such as electrons also exhibited diffraction effects that could only be explained by reference to waves. In each case, both wave and particle aspects are necessary for a complete picture of reality. This "wave-particle duality" lies at the very foundation of quantum mechanics as it developed in the 1920s and beyond [Hoffman 1959; Greenstein 2006]. The sophisticated mathematical formalism of quantum mechanics provides accurate calculations of all physical properties, both particle and wavelike, but a clear physical picture that resolves these apparent paradoxes has been lacking. Indeed, such a picture is often claimed to be unattainable, or even irrelevant to the understanding of quantum mechanics.

Notwithstanding these conventional difficulties, the present paper presents a simple picture and an outline for a more complete theory that incorporates quantum formalism and provides a clear physical foundation for both wave and particle properties of both matter and radiation. The picture is based on real rotating vector fields, which carry energy, momentum, and angular momentum, and provide a local phase reference. Self-interaction of these "rotating spin fields" (RSF) leads to phase coherence, and quantization of spin leads to the condensation of these fields into discrete (but extended) "particles", with energy and momentum related to wave properties by the Einstein-de Broglie relations $E=\hbar\omega$ and $p=\hbar k$ [Kadin, 2005]. This also addresses the quantum measurement question, by providing a dynamical process for transitions between states.

The guiding analogy in understanding this picture is the phenomenon of magnetic domains in a ferromagnet. A magnetic material may be viewed as a field of atomic-scale magnetic moments, which interact with each other via electronic exchange interactions [Kittel, 2004]. These interactions tend to produce coherent alignment resulting in long-range magnetic order below a characteristic transition temperature. Ferromagnetism corresponds to parallel alignment of these magnetic moments, although other coherent relationships are also possible.

If one starts with a magnetic material at high temperature and cools down below the critical temperature, the material will spontaneously break up into an array of domains, each one exhibiting internal ferromagnetic order. However, the coherence is not maintained from one domain to the next. Each single domain can act as an independent "particle" with a macroscopic magnetic moment. For example, it can rotate coherently in response to an external magnetic field. As long as it remains coherent, the internal interactions that maintain this coherence are not visible. But under certain circumstances one can have a rapid transition from one coherent state to another, as for example when a reverse domain nucleates and rapidly spreads. The dynamics of the domain wall motion during the transition reflect the local exchange interactions.

In a similar way, one can think of a fundamental quantum particle (such as an electron or a photon) as a coherent domain in an extended quantum field. The analogy is not precise; the atoms in the magnetic material are normally fixed in space, whereas a quantum field can move freely. Nevertheless, this Coherent Quantum Domain (CQD) picture suggests that an incoherent transition of a quantum particle between states may be governed by different dynamics than the coherent motion of the particle within a given state. A possible form for self-interaction is suggested later in this paper, although the resulting dynamics are not worked out in detail.

This CQD picture seems to be distinct from any of the alternative interpretations of quantum mechanics that have been discussed in the literature [Greenstein, 2006]. It also seems to be consistent with the standard quantum equations, and not in obvious conflict with any of the major experimental results. Given the simplicity and generality of the picture, a careful examination of its derivation and consequences should yield considerable insights into the foundations of quantum theory.

The remainder of the paper first reviews the concepts of waves and particles in classical physics, and then goes on to discuss how wave-particle duality has been treated in quantum theory. Section IV introduces the concept of rotating spin fields, and shows how coherence and quantization of spin lead to the Einstein-de Broglie relations, the heart of quantum mechanics. Section V shows how a nonlinear self-interaction of these rotating fields can bring about coherence and spin quantization, and describes how the picture of Coherent Quantum Domains permits a clear understanding of both wave and particle aspects of quantum systems. Section VI discusses the implications of this for quantum measurement and entangled states. The paper concludes with a discussion of the requirements for a more complete theory of quantum fields and particles, and some broader implications.

## II. Particles and Waves in Classical Mechanics

A classical particle has mass m and position **r**, together with momentum **p** and energy E. This may also be expressed in terms of the Hamiltonian E=H(**p**,**r**). For a particle moving in a potential V(**r**), H=$p^2$/2m +V(**r**), and the standard relations for the velocity **v**=∂H/∂**p** and the force **F**=d**p**/dt = -∂H/∂**r** yield the usual dynamical equations of motion [Goldstein, 2002]. Such a particle need not be restricted to a point in space. More generally, an extended object is distributed with a center of mass, which represents the position of the equivalent point particle. It may also exhibit internal degrees of freedom, such as rotation or vibration. In particular, a rigid body can rotate about a fixed axis in space at an angular velocity ω, with angular momentum L.

Special relativity places some additional constraints on particles, in addition to preventing them from exceeding the speed of light c. For a particle at rest, there is an absolute total energy E=$mc^2$. One can transform E and **p** to a moving reference frame (with v<c) using the Lorentz transform, giving E = γ$mc^2$ and **p** = γm**v**, where γ=$(1-v^2/c^2)^{-1/2}$. This is consistent with the Lorentz-invariant relation $(E/c)^2-p^2$ = $(mc)^2$, which can also be expressed as the square-modulus of the momentum 4-vector **P** = (E/c, **p**).

Of course, any real classical particle is composed of atoms held together by cohesive forces. For example, in an elastic collision of billiard balls, the precise dynamics during the collision depend on the restoring forces in the structure of the ball. But once the balls are no longer in contact, they act as collections of single particles with no evidence of their internal structure. Only if one smashes the balls together with such violence that they shatter does the composite nature of the balls become fully evident.

A magnetic domain would seem to represent a quite different kind of particle, since it is generally fixed in space. But here too, one has a very large number of atomic-level spins that are bound together to form a single unitary object. The total magnetic moment may be free to rotate under the influence of an external magnetic field, in such a way that the internal cohesive forces that maintain this as a single particle are virtually invisible.

There are various kinds of classical waves, but the most important are electromagnetic waves, transverse waves that move (in vacuum) at the speed of light c, and derive their properties from Maxwell's equations [Jackson, 1999]. All waves can be constructed from sinusoidal components cos(ωt-**k**·**r**), where ω is the angular frequency and **k** is the wave vector. The phase angle is θ=ωt-**k**·**r**, and the phase velocity is defined by $v_{ph}$ = ω/k. But more important than $v_{ph}$ is the group velocity $v_g$ = dω/dk, which governs propagation of energy with the wave. For transverse electromagnetic (TEM) waves in free space, ω=ck, so that both $v_{ph}$ and $v_g$ equal c. Furthermore, electromagnetic waves are Lorentz covariant, so that (ω/c, **k**) is properly a 4-vector (with modulus 0), and θ is also a Lorentz invariant (as the inner product of two 4-vectors).

A coherent wave has ω and **k** uniform in time and space, and may be produced by a coherent source, either classical or quantum in nature. Coherent waves lead to interference effects, such as standing waves, with characteristic nodes and antinodes in wave intensity. It is worth noting that a standing wave requires a particular reference frame; if one transforms a standing wave to a different reference frame, the component waves will be Doppler shifted so that they are no longer coherent with each other, and no interference effects are evident.

Waves generally carry both energy and momentum distributed through the wave, proportionally to the intensity of the wave, which goes as the square of the amplitude. For a TEM wave in free space with electrical field amplitude $E_0$, the energy density is $\mathcal{E} = \varepsilon_0 E_0^2/2$, and the momentum density is $\mathcal{P} = \mathcal{E}/c$.

A general TEM wave propagating in the z-direction is typically represented as a sum of components linearly polarized in the x- and y-directions. An equivalent basis set consists of circularly polarized (CP) waves, composed of the sum of linearly polarized components with 90º phase lag beween them, with either left or right circular polarization. It is notable that the CP field vector rotates with fixed amplitude and constant angular velocity, rather than oscillating sinusoidally. While it is customary to represent any sine wave as the real part of a complex exponential, rotation of a vector field about a fixed axis is isomorphic to rotation of a phasor in the complex plane.

An important aspect of classical TEM waves is that a CP wave carries not only linear momentum, but also angular momentum about the propagation direction, distributed through the wave. Furthermore, a standard result from Maxwell's equations [Jackson, 1999; Ohanian, 1986] shows that the angular momentum density $\mathcal{S}$ can be related to the energy and momentum densities by the following relations: $\mathcal{E} = \mathcal{S}\omega$, $\mathcal{P} = \mathcal{S}k$. This will turn out to have particular importance in the extension to quantum mechanics.

Waves are often represented in terms of plane waves with uniform amplitude throughout space. But more realistically, a wave is commonly concentrated in a particular region in space. It may be confined as a guided wave by boundary conditions, or it may exist as a traveling "wave packet" localized in a particular region in space with gradually reducing intensity. Like a particle, a wave packet has a total energy and momentum, and even a position, a center of energy analogous to the center of mass of a distributed solid body. However, unlike a particle, the size of a wave packet may not be smaller than about half a wavelength, due to the mathematical properties of Fourier transforms. (This is also the basis for the Heisenberg uncertainty principle in quantum mechanics, but it is really an aspect of classical waves.) Also unlike a particle, the shape and extent of a wave packet are not generally invariant as it moves through space. A soliton is a wave packet that is more like a classical particle, propagating without dispersion, but this can be present only in special cases in a nonlinear or active medium.

Although wave packets may exhibit concentrations in space similar to that in a particle, the phase angle of a wave is a degree of freedom that is simply not present in classical particles. This gives rise to a range of effects such as interference, diffraction, and standing waves, that are unique hallmarks of wavelike behavior.

## III. Wave-Particle Duality in Quantum Mechanics

Quantum systems exhibit the full range of wavelike behavior, but also require that energy is quantized in multiples of $\hbar\omega$ (for a photon) or that charge is quantized in multiples of e (for an electron). It is difficult to see how to combine both the particle and wave properties in a single physical picture, and this presented a dilemma to the early physicists who were developing the theory of quantum mechanics in the 1920's.

For example, a clear indication of particle properties in photons was provided by the Compton effect (1923), whereby an x-ray photon colliding with a free electron transfers some of its energy and momentum to the electron, in accordance with particle conservation laws. The scattered photon has reduced its frequency, but maintained its cohesion as a single particle. In contrast, a classical TEM wave will cause an electron to reradiate TEM waves components in all directions, at the same frequency but with a phase shift. These two pictures seem incompatible.

Louis de Broglie is well known for his pioneering proposal of matter waves (with what is now known as the de Broglie wavelength $\lambda=h/p$), but he also made a serious early attempt to combine wave and particle properties, both for photons and massive particles [de Broglie, 1960; Hoffman, 1959]. He suggested that in either case, one has both a point particle and an extended wave, both consistent with special relativity. In particular, consider a massive point particle in its rest frame. Assume that this is accompanied by an oscillation at $\omega=E/\hbar$, simultaneously in space in the region around the point particle. If one Lorentz-transforms to another reference frame, this relation can only be consistent if $E=\gamma mc^2$. Furthermore, the oscillations are no longer simultaneous, and using the Lorentz transform must form a wave with wavelength $\lambda=h/p$. Stating this in another way, the phase factor $\omega t-\mathbf{k}\cdot\mathbf{r}$ is a Lorentz invariant if and only if $(\omega/c, \mathbf{k}) = (E/c, \mathbf{p})/\hbar$ is properly a 4-vector. De Broglie initially proposed that the same picture might also apply to a photon with a very small mass, but it is consistent even for m=0.

Another important observation by de Broglie was that the phase velocity of a matter wave in its rest frame is infinite, since $\omega$ is large and k=0. This might seem to violate special relativity, except that it is really the group velocity $v_g = d\omega/dk$ that is relevant. And in this case, $v_g = dE/dp$, which from the classical Hamiltonian equations is always equal to the velocity of the particle (which of course is zero in its rest frame). So the wave packet is automatically constrained to move together with its point particle, if the Einstein-de Broglie conditions for an appropriate relativistic particle are used.

This combined picture of a point particle (providing the energy) and a wave (supplying the phase factor) requires a set of equations for each of these, with a "double solution" of a point singularity and a distributed wave. This was found to be difficult to develop in detail, and was not actively pursued by the major researchers (Heisenberg, Schrödinger, Born, Pauli, etc.) developing quantum theory in Germany. Much later, a similar picture of a "pilot wave" that guides the motion of a point particle was developed further by Bohm, but this never became an established approach [Greenstein, 2006].

Despite the deep insights into the relativistic basis of quantum waves, de Broglie's approach seems to have been largely forgotten. The only aspect that was pursued by Erwin Schrödinger and others is the mapping between the Hamiltonian equation v = dH/dp and the wave relation $v_g$ = dω/dk. If one assumes that E=ℏω, the rest of the (non-relativistic) Schrödinger wave equation follows directly.

However, this does not address the issue of how to incorporate particles into a wave picture. Max Born suggested that the intensity of quantum waves should be viewed as the probability that a point particle is to be found in a particular location within the wave when a "particle measurement" is performed, and this formed the basis for what became the standard "Copenhagen" interpretation of quantum mechanics, promoted by Niels Bohr and others. This interpretation also suggested that the quantum wave was not itself a real physical oscillation like a sinusoidal electromagnetic wave, but rather an abstract complex wave ~exp(iθ). This statistical interpretation was highly unsatisfying to many of the key physicists of the time, including Albert Einstein, Louis de Broglie, and even Schrödinger himself [Hoffman, 1959]. Nevertheless, the very success of quantum theory as a calculational tool for atomic physics tended to sideline these and other critics.

More generally, the standard interpretation of quantum mechanics proposes that one obtains particle parameters when one carries out a particle measurement, and wave properties from a wave measurement. But the theory does not describe the detailed dynamics for the transition of the quantum wave in such a measurement (sometimes called the "collapse of the wave function"), just the statistical distribution of the final results. It is also commonly stated that the quantum wave is not a real physical wave at all, but rather just an expression of limited information about a quantum system. This would seem to represent an incomplete description of physical reality, as discussed by Einstein and others [Einstein, 1935].

## IV.   Rotating Spin Fields and the Einstein-de Broglie Relations

The intrinsic spin of the electron was discovered early in the development of quantum theory (1925), but it was never really treated as a fundamental aspect, perhaps because its universality was not recognized at the time. We show below that identifying rotating vector fields carrying spin with the quantum wave leads to a picture that derives particle properties without the need for any point particles.

As discussed earlier, a classical circularly polarized wave consists of a rotating vector field of fixed amplitude. Furthermore, a CP TEM wave carries angular momentum as well as linear momentum and energy, related for a coherent traveling wave by the equations $\mathcal{E} = \mathcal{S}\omega$, $\mathcal{P} = \mathcal{S}k$. If one now asserts (correctly) that a photon wave packet corresponds to a total spin S=ℏ, then integrating these equations over the volume of the wave packet directly yields the Einstein-de Broglie relations E= ℏω and **p**= ℏ**k**. These relations form the heart of quantum mechanics, and provide the essential link between wave properties (ω,**k**) and particle properties (E,**p**). So an assumption of quantized spin seems to convert a classical wave to a quantum wave-particle. Furthermore, this rotating vector field is naturally represented mathematically by a complex scalar wave function Ψ = |Ψ|exp(iϕ).

This relationship between rotating vector fields and spin need not be limited to photons. For a spin-1/2 particle such as an electron, there is no obvious set of classical wave equations analogous to Maxwell's equations. Nevertheless, let us assume a massive field in its rest frame, described by a vector field coherently rotating at frequency $\omega = mc^2/\hbar$, with distributed energy, momentum, and spin. Let us modify the earlier relations to yield $\mathcal{E} = 2\mathcal{S}\omega$, $\mathcal{P} = 2\mathcal{S}k$ (the factor of 2 is needed for consistency with spin-1/2). Then, if the total spin of the electron is quantized at $\hbar/2$, one again obtains the usual Einstein-de Broglie relations, with $E=mc^2$ in the rest frame, as required. One can also Lorentz-transform to any other reference frame, so that the spin axis is independent of the velocity. Furthermore, this rotating vector field maps onto a complex scalar field that in the nonrelativistic limit satisfies the time-dependent Schrödinger equation (with an $mc^2$ offset for the energy):

$$i\hbar\, \partial\Psi/\partial t = [mc^2 + V(\mathbf{r}) - (\hbar^2/2m)\nabla^2]\Psi$$

This is quite remarkable, and suggests that quantized spin in a distributed rotating wave provides the physical and logical basis for quantum mechanics [Kadin, 2005]. This is quite different from the typical historical approach, where $E= \hbar\omega$ is generally taken as the primary assumption, or the axiomatic approach, where the Schrödinger equation is simply written down. Here, spin is not an intrinsic quantity of a point particle; rather, it is a distributed angular momentum associated with a rotating spin field, or RSF.

If this RSF picture is correct, this implies that all fundamental quantum particles have non-zero quantized spin. And indeed, quarks, electrons, and neutrinos all have spin-1/2, while the photon, gluon, and the weak force vector bosons have spin-1 [Coughlan, 1991]. This fact was not known early in the 20[th] century when quantum mechanics was being developed, and the fundamental quantum wave was believed to be independent of spin. On the other hand, the prevailing models of the weak interaction include a Higgs boson which has not yet been directly observed, but which is believed to be a scalar, spin-0 particle. If such a spin-0 fundamental quantum particle were indeed observed, that would seem to conflict with the present picture. There are, of course, composite quantum particles that have zero spin, such as mesons (composed of quark-anti-quark pairs), alpha particles, and atoms. But the rotating vector fields of the fundamental components are still present and contributing to the coherence of the composite particle, even if their total angular momentum cancels out.

Of course, there already is an established relativistic theory of electrons including spin, namely the Dirac equation [Merzbacher, 1997]. Indeed, the Dirac equation yields an electron that in its rest frame has a phase oscillation at $\omega = mc^2/\hbar$. Furthermore, it has been pointed out that the Dirac equation corresponds to circular energy flow and distributed angular momentum [Ohanian, 1986], which seems to be generally consistent with the simple picture proposed here. Finally, the fundamental physical basis of the Dirac equation is rather obscure, and might be clarified by reference to the present RSF picture.

## V. Coherent Quantum Domains

The arguments in the previous section show that the Einstein-de Broglie relations, which couple wave and particle properties, follow logically from classical wave equations for rotating spin fields based on only two additional assumptions: (1) Wave coherence, and (2) Quantization of spin. A complete theory must explain the basis for these two assumptions, as well as the distinction between fermions and bosons.

Coherence is generally assumed in quantum mechanics as if it were so obvious that it need not be mentioned. However, one can certainly have a classical wave that exists in an arbitary incoherent superposition of harmonic waves. In contrast, quantum states in atoms always seem to exist only in single-frequency energy eigenstates, except during brief transitions between such eigenstates. If one thinks of a quantum wave as consisting of a continuous distribution of independent oscillators (or spinners), then it is natural to think of a local self-interaction that tends to maintain synchronization among these oscillators. This would be analogous to local exchange interactions within a magnetic domain that act to maintain coherence.

One possible way to construct such a self-interaction starts by considering how electromagnetic waves interact with electrons, in both classical and quantum mechanics. The electromagnetic potentials V and **A** (which together form a 4-vector) add to the energy and momentum, so that $E \rightarrow E+eV$ and $\mathbf{p} \rightarrow \mathbf{p}+e\mathbf{A}$, or equivalently for the quantum case $\omega \rightarrow \omega + eV/\hbar$ and $\mathbf{k} \rightarrow \mathbf{k}+e\mathbf{A}/\hbar$ [see e.g., Merzbacher, 1997]. Typically the electromagnetic wave has its own frequency and wavevector $\omega_{em}$ and $\mathbf{k}_{em}$. Within a wave picture, one can think of the electromagnetic wave as modulating the electron wave. As in classical modulation theory, this nonlinear interaction generates sum and difference frequencies $\omega \pm \omega_{em}$ (and similarly for **k**) that correspond in the particle picture to absorption and emission of a photon. In this way, the particle properties of conservation of energy and momentum may derive from wave properties of spacetime modulation.

If the effects of a photon on an electron can be described in terms of an electromagnetic 4-potential (V, **A**), then one should equally be able to describe the back-action of an electron wave on a photon in terms of an effective *electron* 4-potential ($V_e$, $\mathbf{A}_e$), related to the rotating spin field of the electron in a similar way to that for the electromagnetic case. The modulation produced by the electron potential on an x-ray photon yields the transfer of energy and momentum in Compton scattering, for example.

If an electron wave can modulate an electromagnetic wave, might each type of wave also modulate itself? This would correspond, for example, to $\omega \rightarrow \omega + gV_0\exp(i\omega t)$, for some self-coupling strength g. This self-modulation, in turn, would tend to maintain synchronization and coherence across the wave. One might expect a strong self-interaction, so that the quantum wave would maintain coherence in almost all circumstances. It is suggested here (although by no means proven) that while coherence is maintained, the overall wave dynamics would be described by the usual linear quantum equations.

A self-interaction of this sort is similar to frequency modulation (FM) interactions that are well known in systems of nonlinear oscillators, e.g., Josephson junctions. In these systems, an external oscillator will tend to "phase-lock" a given Josephson junction with the same frequency and phase [Kadin, 1999]. Multiple interacting Josephson junctions will tend to synchronize with each other.

This self-interaction provides for an unusual kind of cohesion of a quantum particle. There is a rigidity associated with maintaining coherence, but not with a particular density or distribution of the wave. So the wave can spread out over a larger region of space without weakening its coherence, as long as it maintains spatial continuity.

It is unclear whether such a self-interaction can by itself also account for quantization of spin. Spin quantization is needed to assure that when a distributed quantum wave divides or scatters, each piece has exactly an integral number of particle spins, without fractions permitted. It is well known that quantization of orbital angular momentum (in units of $\hbar$) is topological in nature, following from the de Broglie wavelength around a multiply-connected path. But it is difficult to see how one can derive a topological basis for quantized spin itself, at least without additional spatial dimensions. Of course, the conventional quantum picture does not explain quantized spin either, although it may seem less problematical for point particles.

A more complete picture also needs to account for the distinct characteristics of fermions and bosons. For spin-1/2 fermions, coherence is maintained in a quantum wave only up to a total spin of $\hbar/2$. For larger values of total spin, quantum domains must either move to non-overlapping regions of space, or alternatively shift to wave modes with different resonant frequencies. In contrast, for bosons, one can maintain coherence for larger spins, as long as the total spin is quantized in units of $\hbar$. So one can obtain macroscopic coherence for boson fields (such as electromagnetic waves), but not for fermion fields.

Finally, note that fermion fields of opposite spin can occupy the same quantum state in the same region. Within the picture of rotating spin fields, opposite spin corresponds to rotation about the same axis in the opposite sense. In fact, it is well known that two circularly polarized waves of opposite polarity (and identical amplitudes) give rise to a linearly polarized wave. So a pair of electrons of opposite spin (such as in the ground state of an atom) can be viewed as consisting of coherent linearly polarized fields.

### VI. Implications for Quantum Measurement Theory

The Coherent Quantum Domain picture has important implications for the theory of quantum measurement, for which conventional quantum theory is rather incomplete. Consider the classic two-slit interferometer, which gives rise to interference effects both for single photons and single electrons [Greenstein, 2006]. Clearly, there is a wave going through both slits, and interfering on the other side. But the detector that measures the interference pattern itself typically detects localized "point particles". How does the quantum state change from a wave when it passes through the slits to a particle when it hits the detector? Conventional theory does not have a clear explanation for this apparent "collapse of the wave function".

Within the CQD picture, one can imagine that a photon is incident on an extended detection screen that contains a large number of atomic-scale detectors. An extended photon wave begins to interact with many of these atoms, but given the unitary nature of the photon, in the end it must deposit its entire energy in only one of these. The interactions may be deterministic in principle, but pseudo-random in practice, due to incoherent fluctuations of the various detector atoms. One can think of such an interaction as locally modulating the frequency of the photon wave, tending to cause it to lose coherence with the rest of the wave. If the interaction starts to favor a particular atom, the coherent self-interaction will tend to pull the rest of the photon wavefunction along with it, or else repel it back to the main body of the wave, in order to maintain coherence. The measurement may be said to occur when the transition from an extended wave, to one that has been absorbed on the atomic scale, is complete.

This also suggests that a quantum measurement occurs via a continuous dynamical "collapse of the wavefunction" over a finite interval of time, presumably consistent with special relativity. Since the wavefunction may spread out over macroscopic distances, as in the interference experiment described above, this time may be quite significant. For example, the collapse of a wavefunction spread over 30 cm (1 foot) would require 1 ns (at the speed of light), a time interval that is easily measured in modern experiments. Remarkably, it is not clear that critical experiments have been carried out to measure the time delay associated with quantum measurements. The key problem is that although the completion of a measurement is well-defined by energy transfer to another system (such as a detector), it may be difficult to establish the timing for the initiation of the measurement. Particle emission processes are usually quasi-random, and one does not know that a particle is present until after it has transferred energy to a detector.

One can also consider another classic measurement problem, that of radioactive decay of an atomic nucleus. For example, one may have an alpha particle trapped inside a nucleus, with a large energy required for it to escape. This is a tunneling problem, with a weak tail of the nuclear wave function corresponding to a small probability of escape. Within the conventional picture, this can only be examined statistically. If one considers that the alpha particle is in a coherent state which is partly outside the atom, one may be led to believe that the atom is in a coherent superposition of decayed and un-decayed states. This, in turn, would seem to lead to the "Schrödinger cat problem", where a classical macroscopic system would also seem to be in a similar coherent quantum superposition [Hoffman, 1959, Greenstein 2006]. This is clearly unacceptable, and Schrödinger himself proposed this thought-experiment only to point out shortcomings in the conventional understanding of quantum mechanics.

From the CQD viewpoint, however, a more consistent picture arises. The tail of the wave function starts to escape from the nucleus, and then begins to interact with other localized atoms and systems nearby. In most cases, the weak tail will not be able to "wag the dog", and coherence can be maintained only by reabsorbing the tail part of the wave back into the main body (and possibly even back into the nucleus). However, occasionally (in a pseudo-random sense) the external interaction may be phased just right and with sufficient strength to permit the wavefunction to transfer its energy and momentum to an external system, and the decay has occurred. Note that this does not require a macroscopic measurement instrument or an observer, just an inelastic interaction that changes the energy and hence produces decoherence.

Finally, let us consider the coupled decay of two particles, leading to entangled quantum states [Greenstein, 2006; Zeilinger, 2005]. These provide the basis for what is sometimes called the EPR paradox, after Einstein, Podolsky, and Rosen [Einstein, 1935]. In one form of this paradox, consider the decay of an atomic state that initially has S=0. This is conventionally analyzed in terms of two emitted point photons (for example) going in opposite directions, carrying opposite (coupled) spins. The predictions of quantum theory indicate that the quantum wavefunctions of the two particles are "entangled" (i.e., correlated) until there is a measurement on one of the particles that forces it into a particular state. Then the other particle "immediately" goes into a complementary state. So a measurement on one quantum particle would seem to change the physical properties of another particle, some distance away. This apparent non-locality has been verified by many experiments [Zeilinger, 2005], but would seem to defy both physical intuition and special relativity (and even causality).

However, the CQD picture suggests a possible resolution of this paradox. Consider, for example, the correlated S=0 two-photon state described above. By symmetry, this would correspond to an extended spherically symmetric coherent wavefunction, rather than two separated point particles as in the conventional picture. An interaction of this combined wavefunction with an external system will initiate a quantum measurement process, which will cause both particle wavefunctions to dynamically collapse into complementary localized wavefunctions, consistent with conservation laws. Although only one of the two particles has been detected, the other one has also collapsed, in a similar time period. It is useful to contrast this picture with the common paradoxical version, where detection of point particle #1 somehow sends a message through space to the correlated point particle #2 telling it what state to go into. Within the present picture, the detection of particle #1 indicates the completion of the decorrelation process of the coupled extended wavefunction into two single-particle localized wavefunctions. No paradox is evident. It would be interesting to carefully re-examine recent experimental measurements of entangled states to see if the observations can be consistent with this CQD picture.

## VII. Discussion and Conclusions

Quantum mechanics is virtually unique in the history of physics in representing a successful and widely accepted theory whose fundamental interpretation is still in question 80 years after the theory first appeared. This should indicate that (1) the formalism of quantum mechanics is essentially correct, and (2) the foundations of quantum mechanics have been seriously misunderstood. The present paper attempts to rebuild quantum mechanics based on a new set of foundations, in a way that seems to avoid the standard paradoxes of quantum interpretation.

The major difficulty with the standard approach has been the focus on point particles, which are an idealization even in classical mechanics. A more natural focus is on waves, since the formalism is built around quantum wavefunctions. But the particle aspects are also important, since quantization of energy provided the first indications of inadequacies of classical wave models. The key question then is how to derive particle properties from a wave.

The present proposal is based on two central concepts: Rotating Spin Fields (RSF), and Coherent Quantum Domains (CQD). According to the first concept, all fundamental quantum waves are built around intrinsically rotating vector fields that carry angular momentum, and which map onto the complex scalar quantum wavefunction. According to the second concept, these waves self-organize into coherent domains with quantized total spin, which represent quantum particles. Consistency with special relativity yields the Einstein-de Broglie relations, which connect the local wave properties to the global particle properties. Point particles are not needed, nor is a statistical interpretation of the wavefunction.

In order to obtain a more complete quantum theory, one also needs to understand how these coherent quantum domains interact and undergo transitions from one state to another. Some sort of local interaction is needed to maintain cohesion of any particle. Based on the model of electromagnetic potentials modulating the $\omega$ and **k** of an electron wave, it is suggested that self-modulation by similar potentials might serve to synchronize local field rotations, thus maintaining coherence across the quantum wavefunction. But this also provides a dynamic mechanism for transitions between coherent states, which may constitute a model for quantum measurement. It remains to be shown that such a model is both internally consistent and compatible with experimental observations.

Even if the proposals in the present paper turn out to be incorrect, this analysis should help to stimulate further research into the foundations of quantum mechanics. Both for quantum practitioners and for students learning the field for the first time, such a re-examination is long overdue. Furthermore, recent work in quantum communication and quantum computation indicate real possibilities for practical applications of coherent quantum waves on macroscopic scales. If a simpler, more intuitive understanding of quantum theory free of paradoxes can be developed, that can only help promote these new research areas.

# References

G.D. Coughlan and J.E. Dodd, 1991, *The Ideas of Particle Physics: An Introduction for Scientists*, 2nd ed. (Cambridge U. Press, Cambridge, UK).

L. de Broglie, 1960, *Non-Linear Wave Mechanics: A Causal Interpretation* (Elsevier, Amsterdam, 1960)

A. Einstein, B. Podolsky, and N. Rosen, 1935, "Can Quantum-Mechanical Description of Physical Reality Be Considered Complete?", *Phys. Rev.* **47,** 777.

H. Goldstein, *et al*., 2002, *Classical Mechanics*, 3rd ed. (Addison Wesley, Reading, Mass.).

G. Greenstein and A.G. Zajonc, 2006, *The Quantum Challenge: Modern Research on the Foundations of Quantum Mechanics*, 2nd ed. (Jones & Bartlett, Sudbury, Mass.)

B. Hoffman, 1959, *The Strange Story of the Quantum* (Dover Publications, New York).

J.D. Jackson, 1999, *Classical Electrodynamics*, 3rd ed., p. 350 (Wiley, NY)

A.M. Kadin, 2005, "Circular Polarization and Quantum Spin: A Unified Real-Space Picture of Photons and Electrons," ArXiv Quantum Physics preprint
 http://www.arxiv.org/quant-ph/0508064 .

A.M. Kadin, 1999, *Introduction to Superconducting Circuits* (Wiley, New York).

C. Kittel, 2004, *Introduction to Solid State Physics,* 8th ed. (Wiley, New York).

E. Merzbacher, 1997, *Quantum Mechanics*, 3rd ed. (Wiley, New York).

H.C. Ohanian, 1986, "What is Spin?", *Am. J. Phys.* **54**, 500.

A. Zeilinger, *et al*., 2005, "Happy Centenary, Photon", *Nature* **433**, 230.

---

[*] Email  <amkadin@alumni.princeton.edu>